\documentclass[notoc,12pt]{JHEP3}

\usepackage{amsmath,amssymb,euscript,array,cite}

\input{epsf}

\usepackage{epsfig}

\def\N{{\cal N}}

\def\Tr{{\rm Tr}\,}

\def\det{{\rm det}}
\def\SU{\text{SU}}
\def\U{\text{U}}

\def\Dbarslash{\,\,{\raise.15ex\hbox{/}\mkern-12mu {\bar\D}}}
\def\Dslash{\,\,{\raise.15ex\hbox{/}\mkern-12mu \D}}
\def\delslash{\,\,{\raise.15ex\hbox{/}\mkern-9mu \partial}}
\def\delbarslash{\,\,{\raise.15ex\hbox{/}\mkern-9mu {\bar\partial}}}

\newcommand{\PD}[2]{\frac{\partial #1}{\partial #2}}

\newcommand{\EQ}[1]{\begin{equation} #1 \end{equation}}

\newcommand{\SP}[1]{\begin{equation}\begin{split} #1 \end{split}\end{equation}}

\title{Critical Points of 
Glueball Superpotentials and Equilibria of Integrable Systems}
\author{Timothy J.~Hollowood\\
Department of Physics, University of Wales Swansea,
Swansea, SA2 8PP, UK\\
E-mail: {\tt t.hollowood@swan.ac.uk}}
\preprint{SWAT-}
\abstract{We compare the matrix model and integrable system
approaches to calculating the exact vacuum structure of general $\N=1$
deformations of either the basic $\N=2$ theory or its generalization
with a massive adjoint hypermultiplet, the $\N=2^*$ theory. We show
that there is a one-to-one correspondence between 
arbitrary critical points of the Dijkgraaf-Vafa
glueball superpotential and equilibrium configurations of
the associated integrable system. The latter being either 
the periodic Toda chain, for $\N=2$, or the elliptic Calogero-Moser
system, for $\N=2^*$. We show in both cases that the glueball 
superpotential at the crtical point
equals the associated Hamiltonian. 
Our discussion includes an analysis of the vacuum 
structure of the $\N=1^*$ theory with an arbitrary 
tree-level superpotential for one of the adjoint chiral fields. 
}  

\begin{document}

\section{Introduction}

There are a number of ways of investigating the vacuum structure of an
$\N=1$ supersymmetric gauge theory. In
this paper we have in mind two such theories, 
the basic $\N=2$ theory and also its
generalization with a massive adjoint hypermultiplet (known
as the $\N=2^*$ theory), both with gauge group $\U(N)$, deformed by
an arbitrary superpotential for the adjoint chiral multiplet. We will
focus on two techniques,
based on integrable systems \cite{nick,us1,us2,DS} and
on matrix models \cite{DV1,DV2,DV3,DV4}. The other closely related 
approach involves formulating the problem in terms of Seiberg-Witten theory 
\cite{Cachazo:2001jy,Cachazo:2002pr}. 

In the integrable system approach the vacua are determined
by the extrema of the conserved quantity associated to the 
$\N=1$ deformation. In other words,
the vacua correspond to the equilibria of the associated flow. On
the matrix model side the vacua are determined by the
Dijkgraaf-Vafa glueball superpotential. In this note we show that
there is a one-to-one correspondence between these approaches and at
an equilibrium point the spectral curve of the integrable system is
equal to the auxiliary Riemann surface of the matrix model, extending
the results of \cite{mm2} for the massive vacua 
of the $\N=1^*$ theory. For these special vacua, the
equilibria are stationary with respect to all of the flows of the
integrable system signalled by the fact that 
the spectral curve degenerates all the way to genus one.
For a general vacuum this will
not be the case and the spectral curve will only partially degenerate.
In order
to complete the proof we show that at a critical point the 
value of the glueball
superpotential equals the Hamiltonian in the integrable system.

The relation
between integrable systems and matrix models has also been
investigated in \cite{Boels:2003fh} for the basic $\N=2$ case. As we shall
see, our approach is rather different, although ultimately must be related.
Our approach has some overlap with the recent
paper \cite{PTZ} which considers, in the context of the $\N=2^*$ theory,
the relation between the Riemann
surfaces of the matrix model and the Seiberg-Witten curve
for the $\N=2^*$ theory constructed by Donagi and Witten \cite{Donagi:1995cf}.

\section{The Basic $\N=2$ Case}

First we describe the integrable system approach (for a general review
of integrable systems see \cite{Olshanetsky:dk}). The Coulomb branch of
the $\N=2$ theory with gauge group $\U(N)$ is identified with the
moduli space ${\cal M}_{\rm int}$ of the 
spectral curve $\Sigma_{\rm int}$ of the periodic Toda chain, or 
$A_{N-1}^{(1)}$ affine Toda, integral system. The spectral curve is
the Riemann surface defined by the equation 
\EQ{
F(x,z)=\det\big(x{\bf1}+L(z)\big)=0\ ,
\label{str}
}
where $L(z)$ is the $N\times N$ Lax matrix which depends on the
canonical variables $\{p_i,q_i\}$ via 
\EQ{
L(z)=\begin{pmatrix}p_1 & e^{q_1-q_2} & 0 & \cdots & z\\
                          1 & p_2 & e^{q_2-q_3} & \cdots & 0\\
                          0 & 1 & p_3 & \cdots & 0 \\
                          \vdots & \vdots & \ddots & \ddots & \vdots
			  \\
                      z^{-1}\Lambda^{2N}e^{q_N-q_1} & 0 & \cdots & 1 &
		      p_N\end{pmatrix}\  .
}
In the present context, the canonical variables are
considered to be complex.
The $N$ independent flows of the system can be written in Lax form
\EQ{
\partial_{t_i}L(z)=[M_i,L(z)]\ ,
}
where $M_i$ are $N\times N$ matrices. It is then clear that quantities
of the form
\EQ{
H={\rm Tr}\,\tilde W(L(z))\ ,
\label{cct}
}
for an arbitrary polynomial function $\tilde W(x)$, are conserved. 
In particular, there
are $N$ independent conserved quantities that can be taken to be
\EQ{
H_i={\rm Tr}\,L^i(z)\ ,\qquad i=1,\ldots,N\ .
}
Note that the $\{H_i\}$ are independent of $z$ up to a additive constant
for $i=N$. Any quantity of the form \eqref{cct} is some function of
the $\{H_i\}$ and will generate a flow $\partial_t$ that is some linear
combination of the basic flows $\partial_{t_i}$. 
Note that the basic conserved quantities
$\{H_i\}$ play the r\^ole of coordinates on  
${\cal M}_{\rm int}$, the moduli space of $\Sigma_{\rm int}$.

The curve \eqref{str} can be written more explicitly as
\EQ{
F(x,z)=\prod_{i=1}^N(x-\tilde a_i)+z+\Lambda^{2N}z^{-1}=0\ ,
}
where the $\tilde a_i=\tilde a_i(H_j)$ are an alternative 
set of coordinates on ${\cal M}_{\rm int}$. 
The curve can be written in
hyperelliptic form by defining
\EQ{
y=2z+\prod_{i=1}^N(x-\tilde a_i)
}
in terms of which it takes the form
\EQ{
y^2=\prod_{i=1}^N(x-\tilde a_i)^2-4\Lambda^{2N}\ .
\label{hef}
}
This has the form of a double-sheeted cover of the complex $x$-plane 
with $N$ cuts joining the sheets. Consequently $\Sigma_{\rm int}$ is a
genus $N-1$ Riemann surface.

The centre-of-mass motion associated to $\sum_ip_i$ and
$\sum_iq_i$ is trivial and we can ignore it. The remaining
$N-1$ conjugate angle variables of the integrable system are naturally
identified with a point $\psi_j$, $j=1\ldots,N-1$, 
in the Jacobian torus of the Riemann surface
$\Sigma_{\rm int}$. The Jacobian torus is defined as follows (for
reference on Riemann surfaces see \cite{FK}). First we
choose a canonical set of 1-cycles on $\Sigma_{\rm int}$, $(A_j,
B_j)$, with intersections $A_j\cdot A_k=B_j\cdot B_k=0$, $A_j\cdot
B_k=\delta_{jk}$, $j,k=1,\ldots,N-1$. Let $\omega_j$ be the associated
set of $N-1$ holomorphic 1-forms (abelian differentials of the 1st
kind) normalized so that
$\oint_{A_j}\omega_k=\delta_{jk}$. The period matrix of $\Sigma_{\rm
  int}$ is the $N-1\times N-1$ matrix with elements
\EQ{
\tau_{jk}=\oint_{B_j}\omega_k\ .
}
The Jacobian torus consists of points $\psi_j\in{\mathbb C}^{N-1}$ with
the identifications
\EQ{
\psi_j\thicksim \psi_j+n_j+\tau_{jk}m_k\ ,\qquad n_j,m_k\in{\mathbb Z}\ .
}

We are interested in the flow generated by the conserved quantity of the form
$H$ in \eqref{cct} 
where $\tilde W(x)$ is some polynomial of degree $n$. Notice that
$\tilde W(x)$ itself will depend on $z$ in such a way that $H$ is
$z$-independent. For any choice of Hamiltonian $H$, the associated dynamics 
is linear in the Jacobian. In
other words, for each Hamiltonian $H$, there is a linear flow
\EQ{
\psi_j(t)=\varpi_j t+\psi_j(0)\ ,
}
where $\varpi_j$ are the angular velocities which just
depends on $\Sigma_{\rm int}$. It
is determined by the unique meromorphic 1-form $\Omega$ on
$\Sigma_{\rm int}$ via
\EQ{
\varpi_j=\oint_{B_j}\Omega
\label{defav}
}
normalized by
\EQ{
\oint_{A_j}\Omega=0
\label{ncd}
}
and which is holomorphic on
$\Sigma_{\rm int}-P_\pm$, where $P_\pm$ are the two points at
$x=\infty$ on the upper and lower sheet. The singularities at
$P_\pm$ are specified uniquely by the conserved quantity $H$ in the
following way. For each polynomial $\tilde W(x)$ in \eqref{cct} there is a
unique polynomial $W(x)$ of the same order, for which
\EQ{
\lim_{P\to P_\pm}\Omega(P)=\pm d\big(W'(x)+{\cal O}(1/x)\big)\ .
\label{asww}
}
Since $W(x)$ is some fixed polynomial and doesn't depend on $z$ by choice, 
it must be the case that $\tilde
W(x)$ depends on $z$ in such a way that the associated Hamiltonian $H$
in \eqref{cct} is $z$-independent. We will make the relation between
$W(x)$ and $\tilde W(x)$ more explicit later.

We now consider how to use the integrable system to
find the vacuum structure of the deformed $\N=2$ theory. In the field
theory, the deformation is achieved by adding a tree-level superpotential
\EQ{
\frac1{g_{YM}^2}{\rm Tr}\,W(\Phi)
\label{bss}
}
to the bare Lagrangian of the $\N=2$ theory. Here, $\Phi$ is the $\N=1$
chiral superfield of the $\N=2$ theory. In the low-energy effective
theory the deformation turns on a potential on the Coulomb branch. One
can approach the vacuum problem directly in the four dimensional
theory by using Seiberg-Witten theory. The Seiberg-Witten curve is the
spectral curve of the integrable system and the Coulomb branch is
parameterized by the conserved quantities $\{H_i\}$.
Vacua which survive breaking to $\N=1$, correspond to special
points on the Coulomb branch where a number dyons become massless and
condense after breaking to $\N=1$ 
\cite{Cachazo:2001jy,Cachazo:2002pr}. 
However, in order to relate the problem directly to the
integrable system it turns out to be more useful to compactify the theory on
a circle to three dimensions \cite{nick,us1,us2}. 
This is because the dimension of the Coulomb branch is then doubled by
the addition of the Wilson lines and dual photons of the unbroken $\U(1)^N$
gauge group. The resulting Coulomb branch of the three-dimensional
theory is identified with the (complexified) 
total phase space of the integrable
system where we have not only the Hamiltonians but also the conjugate
angle variables. 

It has been shown \cite{nick,us1,us2} in the context of the $\N=2^*$ theory,
and its quiver generalizations, that the effect of the $\N=1$ deformation 
can be captured---including all the quantum corrections---in 
the three-dimensional theory by taking the superpotential on the
Coulomb branch of the three-dimensional theory to be the conserved
quantity $H$ in \eqref{cct}.
We expect these facts to be true in the basic $\N=2$ theory as well,
since this theory can be obtained by taking the large mass 
limit of the $\N=2^*$ theory. This philosophy has also been advocated
in \cite{Boels:2003fh}. 

Putting aside the usual caveats, supersymmetric vacua are critical
points of the superpotential and are therefore points in the complexified 
phase space which are stationary under the flow $\partial_t$
generated by the
Hamiltonian $H$. The value of the superpotential, {\it i.e.\/}~$H$, 
at the critical point is then valid in the four-dimensional limit.
So from the integrable system point-of-view, as first pointed out in
\cite{us1}, we have to identify
equilibrium points of the complexified integrable system under
the flow $\partial_t$ generated by $H$. This means that the
vacua correspond to points in the moduli space where the vector of angular
velocities $\varpi_j$ vanishes; in other words, at the equilibrium
point, due to 
\eqref{defav} and \eqref{ncd}, it follows that
\EQ{
\oint_{A_j}\Omega=\oint_{B_j}\Omega=0\ .
}
This implies that there exists a meromorphic function $G$ on
$\Sigma_{\rm int}$ such that
\EQ{
\Omega=dG
}
with singularities at $P_\pm=P$ of the form
\EQ{
\lim_{P\to P_\pm}G(P)=\pm W'(x)+{\cal O}(1/x)\ .
\label{asyc}
}
So to summarize: the supersymmetric vacua correspond to points in the
moduli space ${\cal M}_{\rm int}$ where there exists a 
meromorphic function $G$ on $\Sigma_{\rm int}$ 
with a particular pole structure \eqref{asyc} at
$P_\pm$. Notice from $G$ we can form the meromorphic function
$G+W'(x)$ whose only singularity is a pole at $P_+$. Now we can bring  
the Riemann-Roch Theorem to bear on the question of the existence of
$G$. First of all, $P_+$ is not a Weierstrass point\footnote{The
  Weierstrass points $\{P_i\}$ are 
the finite set of points on a Riemann surface
  for which for each $P_i$ 
there exists a non-trivial meromorphic function with a
  singularity only at $P_i$ with an order less than or equal to the genus.}
of $\Sigma_{\rm int}$,
since the latter is hyperelliptic and the Weierstrass points are
located at the $2N$ branch points. Therefore, the existence of a meromorphic
function with a pole of order $n$ at $P_+$ 
requires that $\Sigma_{\rm int}$ degenerates to a surface of genus
$<n$. In other words, in the hyperelliptic form \eqref{hef}, it must
be that 
\EQ{
y^2=\tilde y^2\prod_{i=1}^{N-n}(x-r_i)^2\ ,
}
where 
\EQ{
\tilde y^2=\prod_{i=1}^{2n}(x-s_i)
}
describes a surface of genus $<n$.
In particular, the meromorphic function $G$ is identified 
with $\tilde y$. The pole structure of $G$ \eqref{asyc} then fixes
$\tilde y$ uniquely to be
\EQ{
\tilde y^2=W'(x)^2+f(x)\ ,
\label{swd}
}
where $f(x)$ is some fixed polynomial of degree $n-1$. Note that at
the equilibrium point where $\Sigma_{\rm mm}$ degenerates to a surface
of lower genus $n-1$, the Jacobian must also degenerate. This signals the
fact that the equilibrium point of $\partial_t$ is also stationary
with respect to other flows. In general if the reduced curve has genus
$g$ then there will be $N-g-1$ stationary flows.

These facts dovetail completely with the matrix model approach
\cite{DV1} and, for that matter, the Seiberg-Witten theory approach
\cite{Cachazo:2001jy,Cachazo:2002pr}. 
Without reviewing the matrix model approach, we simply note
that for an $\N=1$ deformation described by $W(x)$ in \eqref{bss}, 
the solution of the 
matrix model involves a hyperelliptic Riemann surface of the form
\EQ{
\tilde y^2=W'(x)^2+f(x)\ ,
}
where $f(x)$ is a polynomial of degree $n-1$. This curve is manifestly
identical with the spectral curve of the integrable system
at the equilibrium point under the flow associated to $H$. 

To complete the picture, we now prove that the value of the
Hamiltonian $H$ that generates the stationary 
flow is stationary is equal to the
critical value of the Dijkgraaf-Vafa superpotential. 
The fact that we can find the Hamiltonian that generates the flow
described by the abelian differential $\Omega$ rests on our knowledge
of the conjugate action variables the the angle variables
\cite{Krichever:1996ut,Krichever:1997sq,D'Hoker:2002pn} (the elliptic
Calogero-Moser case was first considered in \cite{Gorsky:1994dj}. The
conjugate angle variables $a_j$, $j=1\ldots,N-1$, are given by
integrals of a certain abelian differential of the 3rd kind $\lambda$:
\EQ{
a_j=\oint_{A_j}\lambda\ .
}
$\lambda$ is precisely the ``Seiberg-Witten'' 1-form for the $\U(N)$
theory, as deduced in \cite{Klemm:1994qs,Argyres:1994xh}. 
The defining property of $\lambda$ is that
\EQ{
\oint_{A_k}\PD{}{a_j}\lambda=\delta_{jk}\ .
\label{yut}
}
We will take
\EQ{
\lambda=-\log(y+P)dx\ ,
}
which is the appropriate form when the 
derivative in \eqref{yut} is understood
to be taken at fixed $x$. This is necessary in order
that the $a_j$-derivative can be pulled out of the $x$-integral below. 
Defining 
\EQ{
\pi_j=\PD{}{a_j}\lambda\ ,
}
we see that $\pi_j$ is an abelian differential of the 3rd kind
normalized by
\EQ{
\oint_{A_j}\pi_k=\delta_{jk}
\label{ncda}
}
and with simple poles at $P_\pm$. 
We now apply Riemann's bilinear relation to the abelian
differentials $\pi_j$ and $\Omega$:
\EQ{
\frac1{2\pi i}
\sum_{k=1}^{N-1}\oint_{A_k}\Omega\,\oint_{B_k}\pi_j-\oint_{B_k}\Omega
\,\oint_{A_k}\pi_j={\rm Res}_{P_+}(W'(x)\pi_j)-{\rm
  Res}_{P_-}(W'(x)\pi_j)\ ,
}
where we have used the asymptotic forms for $\Omega$ at $P_\pm$ in
\eqref{asww}. Now we use the normalization conditions \eqref{ncd} and
\eqref{ncda}, along with the definition of the angular velocities
\eqref{defav} and the fact that the contribution from $P_-$ is minus
that at $P_+$, to arrive at
\SP{
\frac1{4\pi i}\varpi_j&=-
{\rm Res}_{P_+}\big(W'(x)\pi_j\big)\\
&=\PD{}{a_j}{\rm Res}_{P_+}\big(W'(x)\log(y(x)+P(x))dx\big)\\
&=-\PD{}{a_j}{\rm Res}_{P_+}\Big(W(x)\frac{P'(x)}{y(x)}dx\Big)
}
where in the last line we integrated by parts. Now since $a_j$ is
canonically conjugate to the angle $\psi_i$, this means that the
Hamiltonian which generates the flow associated to $\Omega$ is
\EQ{
H=-{\rm Res}_{P_+}\Big(W(x)\frac{P'(x)}{y(x)}dx\Big)\ .
\label{hamr}
}
It is straightforward to show that if $W(x)$ is a polynomial of order
less than $N$ then
\EQ{
H={\rm Tr}\,W(L(z))\ ,
}
where $L(z)$ is the Lax matrix. So in this case we can identify
$\tilde W(x)$ and $W(x)$. If $W(x)$ has an order $\geq N$ then $\tilde
W(x)$ will be $z$-dependent. For instance, for $W(x)=x^N$
\EQ{
\tilde W(x)=x^N+(-1)^N\big(z+\Lambda^{2N}/z\big)\ .
}
 
What is particularly nice about the result \eqref{hamr} is that at a
critical point it agrees precisely with the critical value of the
Dijkgraaf-Vafa superpotential. In order to see that, one simply
deforms the contour around infinity to a sum of contours around the
cuts on the degenerated surface. We then identify the differential
$P'\,dx/y$ with the resolvent of the field $\Phi$ 
defined in \cite{Cachazo:2002ry,Cachazo:2003yc}
\EQ{
T(x)={\rm Tr}\,\frac{dx}{x-\Phi}\ .
}
The critical value of the superpotential is then
\EQ{
\int_{\rm cuts}W(x)T(x)
}
which is equal to \eqref{hamr}. Hence we find perfect argument between
the two distinct approaches for calculating the vacuum structure.

\section{The $\N=2^*$ Case}

We now apply the same philosophy established in the basic $\N=2$ case to
$\N=1$ deformations of the $\N=2^*$ theory. 

The Coulomb branch of the $\U(N)$ $\N=2^*$ theory 
is identified with the moduli space of the spectral curve $\Sigma_{\rm int}$ 
of the $N$-body elliptic Calogero-Moser integrable
system \cite{Donagi:1995cf,Martinec:1995qn}:
\EQ{
F(v,z)=\det\big(v{\bf 1}+L(z)\big)=0\ ,
}
where the $N\times N$ Lax matrix $L(z)$ has components
\EQ{
L_{ij}(z)=p_i\delta_{ij}+m(1-\delta_{ij})\frac{\sigma(z-q_i+q_j)}
{\sigma(z)\sigma(q_i-q_j)}e^{\zeta(z)(q_i-q_j)}\ .
}
We denote the $N$-dimensional moduli space of the curve by ${\cal
  M}_{\rm int}$. 
The quantity $v$ is a meromorphic function on $\Sigma_{\rm int}$ with $N$
simple poles at the pre-images of $z=0$ with residues
\EQ{
m\big(N-1,-1,-1,\ldots,-1\big)\ .
\label{pst}
}
Notice that one of the points, which we denote $P_0$, is distinguished
by the fact that the residue is $m(N-1)$. In the Type IIA brane
construction of Witten \cite{Witten:1997sc} 
$P_0$ is the position of the NS5-brane.
From a field theory perspective, $m$ is the mass of the adjoint
hypermultiplet and $\tau$ is the bare complexified gauge coupling.

$F(v,z)$ is a polynomial of degree $N$ in $v$ whose coefficients are
elliptic functions on the torus $E_\tau$ with complex structure $\tau$:
\EQ{
F(v,z)=\sum_{i=0}^Nf_i(z)v^i\ ,
}
where
\EQ{
f_i(z+2\pi i)=f_i(z+2\pi i\tau)=f_i(z)\ .
}
The spectral curve $\Sigma_{\rm int}$ 
describes an $N$-sheeted cover of the torus
$E_\tau$ joined by branch cuts to make a higher genus surface. The
number of branch cuts corresponds to the number of zeros of $\partial_v
F(v,z)$. Since the latter is a meromorphic function on $\Sigma_{\rm
  int}$ the degree of its zeros is equal to the degree of its poles. It
follows from \eqref{pst} that $\partial_vF(v,z)$ has $N-1$ simple poles
and a pole of order $N-1$ at the $N$ pre-images of $z=0$. Hence, there
are $2(N-1)$ branch cuts and the Riemann-Hurwitz Theorem gives the
genus of $\Sigma_{\rm int}$ as $N$. We can view the surface as $N$
copies of the torus $E_\tau$ 
glued together by $N-1$ tubes to make a genus $N$ surface.

Any quantity of the form 
\EQ{
H={\rm Tr}\,\tilde W(L(z))\ ,
\label{poo}
}
for a polynomial function $\tilde W(x)$, will be conserved. A basis for
the Hamiltonians, and so a set of coordinates for ${\cal M}_{\rm int}$,
is provided by
\EQ{
H_i={\rm Tr}\,L^i(z)\ , \qquad i=1,\dots,N\ ,
}
where $z$ takes a fixed value $\neq0$.
The $N$ conjugate angle variables are, as before, associated with a
point $\psi)j$ in Jacobian of $\Sigma_{\rm int}$. The dynamics is
linear in the Jacobian \cite{Olshanetsky:dk,K1,Krichever:1994vg} and 
for the flow corresponding to an arbitrary Hamiltonian $H$ in
\eqref{poo}, we have
\EQ{
\psi_j(t)=\varpi_j t+\psi_j(0)\ ,
}
where $\varpi_j$ is the vector of angular velocities associated to
$H$. This quantity is determined by
the unique meromorphic 1-form $\Omega$ on $\Sigma_{\rm int}$ with
\EQ{
\oint_{A_j}\Omega=0\ ,\qquad\varpi_j=\oint_{B_j}\Omega
}
such that it is holomorphic on
$\Sigma_{\rm int}-P_0$ with a given singularity at $P_0$ determined by
$H$. Let $x$ be a coordinate in the neighbourhood
of $P_0$ with $x(P_0)=\infty$, then the singularity of $\Omega$ has the form
\EQ{
\lim_{P\to P_0}\Omega(P)=d\big(U(x)+{\cal O}(1/x)\big)\ ,
\label{oas}
}
where $U(x)$ is a polynomial in $x$ of the same degree as
$\tilde W(x)$. In our philosophy, $U(x)$ will be fixed uniquely by the
$\N=1$ deformation, as we shall see later. This then fixes the
Hamiltonian $H$ and hence the function $\tilde W(x)$. 
In particular, since $U(x)$ is by choice 
$z$-independent, $\tilde W(x)$ must be $z$-dependent in
such a way that $H$ in \eqref{poo} is independent of $z$.

Following the logic of the last section, 
we now use the structure of the integrable system to
find the vacuum structure of the $\N=2^*$ theory deformed to $\N=1^*$. The
deformation is achieved by adding a tree-level superpotential
\EQ{
\frac1{g_{YM}^2}{\rm Tr}\,W(\Phi)
\label{nsdef}
}
to the bare Lagrangian of the $\N=2^*$ theory. Here $\Phi$ is massless
adjoint chiral superfield of the $\N=2^*$ theory. As before we
compactify to three dimensions and identify the Coulomb branch of the
theory with the (complexified) total phase space of the integrable
system.

The deformation \eqref{nsdef} is captured exactly by the
superpotential on the Coulomb branch of three-dimensional theory which
is equal to the Hamiltonian $H$ whose flow was described above. Note
that in general the two polynomial functions $W(x)$ and $\tilde W(x)$
are not equal, however, they are of the same order and are uniquely
related to one another---at 
least implicitly---as we shall see later. Supersymmetric vacua are critical
points of the superpotential and are therefore points in the complexified 
phase space which are stationary under the flow $\partial_t$
generated by the Hamiltonian $H$. Consequently, 
at these point the associated angular velocities must vanish; hence,
\EQ{
\oint_{A_j}\Omega=\oint_{B_j}\Omega=0\ .
}
This implies that there exists a meromorphic function $G$ on
$\Sigma_{\rm int}$ such that
\EQ{
\Omega=dG\ ,
}
with, from \eqref{oas}, a singularity at $P_0$, of the form
\EQ{
\lim_{P\to P_0}G(P)=U(x)+{\cal O}(1/x)\ .
\label{asyu}
}

If $W(x)$ is a polynomial of degree $n$ then so are both $\tilde W(x)$
and $U(x)$. Generically, according
to the Riemann-Roch Theorem, for $G$ to exist, 
$\Sigma_{\rm int}$ must degenerate to a surface of genus
$n-1$. However, if $P_0$ happens to be a Weierstrass point of
$\Sigma_{\rm int}$ then there may exist additional 
vacua where $\Sigma_{\rm int}$ has genus $>n-1$. In particular, $n$
would have to be in the ``non-gap'' sequence at $P_0$.
Unlike in the
hyperelliptic situation described in the last section these exceptional cases
do indeed occur. An example occurs in the $\U(3)$ theory with
the simplest quadratic deformation $W(x)=x^2$.
In this case, there is a vacuum described in \cite{Donagi:1995cf} where the
surface degenerates from genus 3 to 2 but not all the way to genus 1
as would be required if $P_0$ was a generic point on the surface. 
In particular, for the quadratic deformation, as was first pointed 
out in \cite{us1} all the vacua must be described by a degeneration to a
hyperelliptic surface since it is only on these surfaces that there
exists a meromorphic function with a double pole. In particular,
$P_0$, being a Weierstrass point, 
must lie at one of the branch points in the two-sheeted representation.
We will leave a more in-depth 
discussion of these exceptional cases to future work.

Before we move on to describe the matrix model approach to the same
problem, we note that we can map the curve $\Sigma_{\rm int}$ into the
complex plane parameterized by $x$ via
\EQ{
x=iv(z)+im\Big(\zeta(z)-\frac{\zeta(\pi i)z}{\pi i}\Big)\ .
}
Here, $\zeta(z)$ is the Weierstrass $\zeta$-function which is a 
quasi-periodic function on $E_\tau$:
\EQ{
\zeta(z+2\pi i)=\zeta(z)+2\zeta(\pi i)\ ,\qquad\zeta(z+2\pi
i\tau)=\zeta(z)+2\zeta(\pi i\tau)\ .
}
In the $x$-plane, $P_0$ is mapped to $x=\infty$. Note that $x$ is
multi-valued on $\Sigma_{\rm int}$ because, although it is 
periodic around the pre-images $a_i$ of the $a$-cycle of the base
torus $E_\tau$, it
picks up an additive constant $im$ around the pre-images $b_i$ of the
$b$-cycle of the base torus. So restricting $x$ to a single sheet
there are $N$ pairs of cuts $C^-_i=[k_i,l_i]$ and
$C^+_i=[k_i+im,l_i+im]$ which are identified. In this picture the
genus $N$ surface $\Sigma_{\rm int}$ is realized as the complex
$x$-plane with $N$ handles formed by identifying the top/bottom
$C^+_i$ with the bottom/top of $C^-_i$. In general a Riemann surface
of this form has $2N$ complex moduli which we can take to be the
positions of the ends of the lower cuts $\{k_i,l_i\}$. We denote the
moduli space of these surfaces as $\hat{\cal M}$. Clearly the moduli
space of 
$N$-fold covers of the base torus $E_\tau$, ${\cal M}_{\rm int}$, is only
a complex $N$-dimensional subspace of this larger moduli space. 
Note that when the surface $\Sigma_{\rm int}$ degenerates, cuts in the
$x$-plane merge. Note that two pairs of cuts can annihilate in two distinct
ways. Either the bottom cut of one pair merges with the bottom
cut of another pair to leave a single pair of the same kind, or 
the top cut of one pair merges with the bottom cut
of another pair resulting in another pair of cuts now separated by
$2im$ rather than $im$. By iterating this procedure we see that
degenerations of $\Sigma_{\rm int}$ are manifested in the complex
$x$-plane by pairs of cuts which join to form handles  
which can be separated by any integer multiple of $im$.

Before we discuss the matrix model approach let us consider the vacuum
structure from the point-of-view of the tree-level superpotential:
\EQ{
W=\frac1{g_{YM}^2}
{\rm Tr}\big(i\Phi[\Phi^+,\Phi^-]+m\Phi^+\Phi^-+W(\Phi)\big)\ .
}
The case with a quadratic superpotential $W(\Phi)$ was considered
originally in \cite{Donagi:1995cf}, here, we present the
generalization for arbitrary polynomial functions $W(\Phi)$.
At tree-level, the supersymmetric vacua can be found by solving the
$F$-flatness conditions
\EQ{
[\Phi,\Phi^+]=im\Phi^+\ ,\qquad[\Phi,\Phi^-]=-im\Phi^-\ ,\qquad
[\Phi^+,\Phi^-]=iW'(\Phi)
}
modulo complex gauge transformations.
Using the symmetries we can diagonalize $\Phi$ and the build up
solutions from a series of irreducible blocks. In such a block of size $p$
$\{\Phi^\pm,\Phi\}$ have the same non-zero elements 
as $\{J^\pm,J_3\}$ of the irreducible representation of $\SU(2)$ of size
$p$. In particular
\EQ{
\Phi=imJ_3+\lambda{\bf 1}\ .
}
The parameter $\lambda$ is then determined by demanding
\EQ{
{\rm Tr}_{\rm block}\,[\Phi^+,\Phi^-]=0=
i\sum_{j=1}^pW'\big(\lambda+\tfrac{im}2(p+1-2j)\big)\ .
\label{hggt}
}
If $W(x)$ is a polynomial of degree $n$, there are $n-1$ possibilities
for $\lambda$. Hence, a general vacuum corresponds to the data
\EQ{
\{n_j,p_j,\lambda_j,\ j=1,\ldots,g\}\ ,\qquad N=\sum_{j=1}^gn_jp_j
\qquad (p_i\neq p_j\text{ when }\lambda_i=\lambda_j)\ ,
}
where $n_j$ is the number of blocks of size $p_j$ associated to
one of the $n-1$ roots $\lambda_j$ of \eqref{hggt}. 
The unbroken gauge group in
this vacuum is
\EQ{
\prod_{j=1}^g\U(n_j)
}
and in particular the number of abelian factors is $g$. 
Quantum mechanically we expect in the infra-red
that the non-abelian parts of the gauge group confine to leave an
abelian theory with gauge group $\U(1)^g$. Note that the maximal value
of $g$ is $N$, obtained when the potential $W(x)$ is a polynomial of
degree $>N$ and $p_j=n_j=1$, $j=1,\ldots,N$. In this case the unbroken
gauge group is $\U(1)^N$. This is the $\N=2^*$ analogue of the vacuum
considered in \cite{Cachazo:2002pr} which can track the Coulomb branch
of the $\N=2^*$ theory and be used to extract the Donagi-Witten curve from
the matrix model (as recently considered in \cite{PTZ}). 
The minimal value of $g=1$ is obtained when $p_1n_1=N$. In other words
when $p_1$ is an integer divisor of $N$. These are the ``massive''
vacua considered in \cite{nick,mm1,mm2} (note that in the $\U(N)$, as
opposed to the $\SU(N)$ theory, there is an unbroken $\U(1)$ factor and
so strictly speaking the vacua are not massive). 

Now we turn to the matrix model approach to calculating the exact
superpotential. According to Dijkgraaf and Vafa we consider a matrix
model whose matrix variables are associated to the chiral superfields
of the theory and whose action is the $F$-term. In the $\N=2^*\to1^*$
theory this yields a matrix model with a partition function (see
\cite{DV3,mm1,mm2,cft,lsint,5d} for previous work on matrix models and
the $\N=1^*$ theory and its generalizations):
\EQ{
Z=\int [d\Phi^+][d\Phi^-][d\Phi]\exp-
g_s^{-1}\Tr\left(i\Phi[\Phi^+,\Phi^-]+m\Phi^+\Phi^-+W(\Phi)\right)\ .
\label{mm}} 
Since $\Phi^\pm$ appear Gaussian we may integrate them out:
\EQ{Z=\int[d\Phi] \frac{e^{-g_s^{-1}\Tr V(\Phi)}}{{\det(
\text{Adj}_\Phi+im)}}\ .
\label{mm2}
}
In order to avoid confusion, we will suppose that the matrices have
size $\hat N$.
In order to implement the Dijkgraaf-Vafa procedure to the vacua
described above, we need to solve the matrix 
model Eq.~\eqref{mm2} in the large $\hat N$-limit
around the saddle-point corresponding to the classical solution for 
where $\Phi$ takes its
tree-level form with $n_i$ replaced by arbitrary variables $\hat n_i$
which individually tend to infinity. 
As usual in the large-$\hat n_i$ limit the eigenvalues of
$\Phi$ spread out from their classical values along cuts on the complex
eigenvalue $x$-plane. In other words, for each $j=1,\ldots,g$
there will a set of $p_j$ cuts which form a group, each element
of which being the image of the 
lower one under translations
$ikm$, $k=1,\ldots,p_j-1$. Each group is located in the vicinity
of $\lambda_j$. So for each $j=1,\ldots,g$ there are 2 complex parameters
which one think of as the two ends of the lower cut. The density of
eigenvalues $\rho(\phi)$ only has non-zero support along the cuts in the
$x$-plane. Moreover, for a each $j=1,\ldots,g$ the density of
eigenvalues along each of the $p_j$ cuts in the group is equal.
The saddle-point equation in the large-$N$ limit is
most conveniently written in terms of the resolvent function
\EQ{
\omega(x)=\int_{\rm cuts}\frac{\rho(\phi)}{x-\phi}d\phi\ ,
\qquad\int_{\rm cuts}\rho(\phi)d\phi=1\ .
}
The resolvent $\omega(x)$ is an analytic function on the complex
$x$-plane whose only singularities are branch cuts along the cuts
where the eigenvalues are located for which the 
discontinuity across the cut gives the eigenvalue density
\EQ{
\omega(\phi+i\epsilon)-\omega(\phi-i\epsilon)=-2\pi
i\rho(\phi)\ ,
\qquad\phi\in{\rm cuts}\ .
}
In this, and following equations, $\epsilon$ is a suitable
infinitesimal regulator. 
The saddle-point equation expresses the condition of  
zero force on a test eigenvalue in the presence of the 
large-$N$ distribution of eigenvalues along cuts:
\EQ{
\frac{W'(\phi)}S=\omega(\phi+i\epsilon)+\omega(\phi-i\epsilon)
-\omega(\phi+im)-\omega(\phi-im)\ ,\qquad\phi\in{\rm cuts}\ ,
\label{spe}
}
where $S=g_s\hat N$ is the 't~Hooft coupling.
This equation can be re-written in terms of the useful function
\EQ{
G(x)=U(x)+iS(\omega(x+\tfrac{im}2)-\omega(x-\tfrac{im}2))\ ,
\label{defG}
}
where $U(x)$ is a polynomial in $x$ of the same degree as $W(x)$ such that
\EQ{
W'(x)=-iU(x+\tfrac{im}2)+iU(x-\tfrac{im}2)\ .\label{defU}
}
It turns out that $G(x)$ has a somewhat simpler analytic structure
than $\omega(x)$. For a given group of $p_j$ cuts the situation is
described in \cite{mm2}. In taking the difference of the resolvents in
\eqref{defG} most of the cuts cancel to leave only a pair of cuts
associated to each group $[k_j,l_j]$
and $[k_j+imp_j,l_j+imp_j]$. Moreover the saddle-point equations simply
imply a gluing condition between each pair of cuts where the
bottom/top of the lower cut is
identified with the top/bottom of the upper cut. This naturally
defines a Riemann surface of genus $g$ since each pair of cuts once
glued corresponds to one handle. In fact it should not escaped the
reader's notice that this Riemann surface is precisely an 
example of the family of surfaces
$\hat{\cal M}$ defined earlier in the context of the integrable
system. 

However there is more structure since the Riemann surface has to admit
a meromorphic function $G$ with a singularity at the point $P_0$, 
$x(P_0)=\infty$, with the following pole structure
\EQ{
\lim_{P\to P_0}G(P)=U(x)+{\cal O}(1/x)\ .
\label{asyf}
}
So the saddle-point equation of the
matrix model boils down to the existence 
of a genus $g$ Riemann surface $\Sigma_{\rm
  mm}$ in the space $\hat{\cal M}$ 
which admits a meromorphic function $G$ whose only singularity is
at the point $P_0$ with the specific pole structure \eqref{asyf}. 

The question is how many moduli does the surface $\Sigma_{\rm mm}$
have? Potentially there are $2g$ complex moduli as pointed out
above. However, there are additional constraints arising from the
requirement that the function $G$ exists on $\Sigma_{\rm mm}$.
Generically, by the Riemann-Roch theorem, such a meromorphic
function only exist on a surface of genus $g<n$, in which case
there would be $n-g$ non-trivial meromorphic functions with a pole of
this order or less. Therefore prescribing the asymptotic form \eqref{asyf}
amounts to $g$ non-trivial conditions of the surface $\Sigma_{\rm
  mm}$, leaving a $g$ complex dimensional subspace of moduli.
The $g$ moduli of these surfaces can be
described by the $g$ quantities
\EQ{
S_i=-\frac1{2\pi}\oint_{A_i}G(x)dx\ ,
}
where $A_i$ is a contour which encircles the lower cut 
$[k_i,l_j]$ of each pair. Notice that the number of matrix model eigenvalues
associated to each $j=1,\ldots,g$ is equal to $p_jS_j/g_s$ and furthermore
\EQ{
S_j=g_s\hat n_j\ ,\qquad S=g_s\hat N=\sum_{j=1}^g p_jS_j\ .
}
Each of the moduli
$S_j$ will become a field of the Dijkgraaf-Vafa superpotential
identified with the glueball superfield of the unbroken $\U(n_j)$ factor
of the gauge group. 

The second ingredient required to construct the Dijkgraaf-Vafa
superpotential is the variation of the genus zero free energy ${\cal
  F}_0$ with respect to $S_j$. Following \cite{mm2}, this is equal to
\EQ{
\PD{{\cal F}_0}{S_j}=-i\oint_{B_j}G(x)dx\ ,
\label{iio}
}
where $B_j$ is the conjugate cycle to $A_j$, in other
words, goes from the lower cut to the upper cut of a pair. The
quantity \eqref{iio} can be interpreted physically as the variation of
the genus zero free energy of the matrix model in transporting $p_j$
eigenvalues in from infinity and placing one on each of the $p_j$ cuts
in a group (so as to maintain the same density along each of the cuts in
the group). We conjecture that the
generalization of the Dijkgraaf-Vafa glueball superpotential for the vacuum in
question is then
\EQ{
W_{\rm eff}(S_j)=\sum_{j=1}^g\Big(n_j\PD{{\cal F}_0}{S_j}-2\pi ip_j\tau
S_j\Big)\ ,
}
where now the $n_j$ are the physical rather than matrix model
quantities.

A critical point of $W_{\rm eff}$ corresponds to
\EQ{
\sum_{j=1}^gn_j\frac{\partial^2{\cal F}_0}{\partial S_k\partial
  S_j}=2\pi i\tau p_k\ .
}
This equation can be written in a more suggestive way by noticing that 
\EQ{
\omega_j=-\frac1{2\pi}\frac{\partial}{\partial S_j}G(x)dx\ ,
}
are a basis for the $g$ holomorphic 1-forms on $\Sigma_{\rm mm}$ since
the singular part of $G(x)dx$ at $x=\infty$ is manifestly independent of the
moduli $\{S_j\}$. Furthermore, the $\omega_i$ are normalized so that
\EQ{
\oint_{a_j}\omega_k=\delta_{jk}\ .
}
Hence 
\EQ{
\frac{\partial^2{\cal F}_0}{\partial S_k\partial
  S_j}=2\pi i\oint_{B_j}\omega_k=2\pi i\tau_{jk}\ ,
}
where $\tau_{jk}$ is the period matrix of $\Sigma_{\rm
  mm}$. Consequently the critical point equations are 
\EQ{
\sum_{j=1}^gn_j\tau_{jk}=\tau p_k\ .
\label{conc}
}
These equations are precisely the conditions that $\Sigma_{\rm
  mm}$ is an $N$-sheeted covering of the base torus $E_\tau$.
In order to prove this we need to find a map
from $\Sigma_{\rm mm}$ to $E_\tau$ which covers the latter $N$
times. For $P\in\Sigma_{\rm mm}$, the map is simply
\EQ{
z(P)=2\pi i\int_{P'}^P\sum_{j=1}^gn_j\omega_j\quad\text{mod }2\pi
i,2\pi i\tau\ ,
}
where $P'$ is a fixed but arbitrarily chosen point of $\Sigma_{\rm mm}$. Since
$N=\sum_jn_jp_j$ this map covers $E_\tau$ $N$ times. In particular,
$z$ is identified with the same quantity on the integrable system
side. Similar ideas have been expressed in \cite{PTZ} for 
the case when $g=N$,
$n_j=p_j=1$ corresponding to the vacuum with unbroken $\U(1)^N$ symmetry.

Now we compare the integrable system with the matrix model. 
The first point is that the moduli space ${\cal M}_{\rm int}$ and the
union of 
the moduli spaces of the matrix model for different vacua ${\cal
  M}_{\rm mm}$ are both subspaces of the 
moduli space $\hat{\cal M}$ defined earlier. The subspace 
${\cal M}_{\rm int}$ contains the surfaces which are
$N$-fold branched coverings of the torus $E_\tau$ and the subspace
${\cal M}_{\rm mm}$ contains surfaces which admit a
meromorphic function $G$ with a fixed pole structure at $P_0$
determined by the $\N=1$ deformation. 
The critical-point condition in the integrable system approach boils
down to the existence of the meromorphic function 
$G$ while on the matrix model side it boils
down to the constraint of being an $N$-fold cover of 
the torus $E_\tau$. In other
words, the vacua correspond to points of intersection between ${\cal
  M}_{\rm int}$ and the different components in ${\cal M}_{\rm
  mm}$. In this picture we must identify 
the polynomial $U(x)$ on both
sides of the story which then implicitly determines 
the relation between $\tilde W(x)$ and $W(x)$. Notice for the case of
the vacuum with $g=N$ and unbroken $\U(1)^N$ we can, generalizing the
situation in the basic $\N=2$ theory \cite{Cachazo:2002pr}, extract the
Donagi-Witten curve from the matrix model model as was pointed out in
\cite{PTZ}.  

Now we turn to the question of the value of the 
superpotential at the critical points calculated using the two
methods. We start by finding a more explicit expression for the
critical value of the superpotential on the matrix model side by the
following manipulations:
\SP{
W_{\rm eff}&=-i\sum_{j=1}^g\Big(n_j\oint_{B_j}G(x)\,dx-\tau
p_j\oint_{a_j}G(x)\,dx\Big)\\
&=-i\sum_{j=1}^gn_j\Big(\oint_{B_j}G(x)dx-\sum_{k=1}^g\tau_{jk}\oint_{A_k}
G(x)\,dx\Big)\\
&=-2\pi\,{\rm Res}_{P_0}\big(U(x)z\,dx\big)\ .
\label{mmre}
}
In the above, to reach the second line we used the critical-point equations
\eqref{conc} and to reach the final line we applied 
a Riemann bilinear relation, used the fact 
that $dz=\sum_jn_j\omega_j$ and that $G(x)$ can be replaced with
$U(x)$ in the vicinity of $P_0$. A similar
expression was derived in \cite{PTZ}.

Now from the integrable system side. Once again we use the fact that
action variables conjugate to the angle are integrals of the
Seiberg-Witten differential:
\EQ{
a_j=\oint_{A_j}\lambda\ ,\qquad\lambda=v\,dz\ .
}
Since the residue of $\lambda$ at $P_0$ is independent of the moduli
$a_i$, the derivative of $\lambda$ with respect to $a_i$ is the
holomorphic 1-form $\omega_i$. However, with the Seiberg-Witten
differential in the form $v\,dz$ we must be careful to differentiate
at fixed $z$ otherwise singularities arise:
\EQ{
\omega_i=\Big(\PD v{a_i}\Big)_z\,dz
}
It turns out that this is not a conventient form. One can eaily
verify from the fact that $F(v,z;{a_i})=0$ and from the relation
between $x$ and $v$ and $z$, that
\EQ{
\omega_i=\Big(\PD v{a_i}\Big)_z\,dz=-\Big(\PD z{a_i}\Big)_v\,dv=
-\Big(\PD z{a_i}\Big)_x\,dx\ .
}
The final expression here is the one which will be most convenient.
Applying a Riemann bilinear relation to $\Omega$ and
$\omega_i$ one arrives at the following expression
for the angular velocities
\EQ{
\varpi_j=-2\pi i\PD{}{a_j}{\rm Res}_{P_0}\big(U(x)z\,dx\big)\ .
}
Hence the Hamiltonian which degenerates the flow described by the
abelian differential $\Omega$ is
\EQ{
H=-2\pi\,{\rm Res}_{P_0}\big(U(x)z\,dx\big)
\label{intre}
}
which is in perfect agreement with the matrix model result \eqref{mmre}. In a
sequel we shall show how to write \eqref{intre} in terms of the Lax
matrix of the integrable system and hence has a function of the
positions and momenta.

Obviously important questions remain, the most interesting concerning 
the exceptional vacua for which the surface has a genus greater than
or equal to the order of $W(x)$ and for which $P_0$ must be a Weierstrass
point. 

\vspace{2cm}
I would like to thank Harry Braden, Jan de Boer, Nick Dorey, Prem Kumar and 
Kazutoshi Ohta for discussions.

\end{document}